\documentclass[%
reprint,
superscriptaddress,
showkeys,
amsmath,amssymb,
pra,
longbibliography,
floatfix,
nobalancelastpage
]{revtex4-2}

\usepackage{defs}
\usepackage{title-and-authors}

\begin{document}
\mainmanuscript

\begin{abstract}
  We introduce the concept of \emph{spin migration}, i.e.\
  time-dependent spin density, and the absolute smallest example of
  molecular spintronics. We illustrate the concept by \emph{ab initio}
  calculations of the coupled vibronic motion of the electrons and the
  nuclei in the argon dimer cation, in which the vibrational level
  spacing is similar to the spin--orbit splitting of the atomic argon
  cations. We find that the spin density varies on multiple time
  scales, given by the natural time scales of the electronic and
  nuclear wavepackets.
\end{abstract}


\maketitle

\section{Introduction}
Upon sudden removal of an electron from a neutral molecule, the
remaining electrons do not have time to relax, and a non-stationary
electronic wavepacket is formed, resulting in a quasi-periodic
oscillatory motion commonly referred to as \emph{charge migration}
\cite{Breidbach2003,Kuleff2005}. In the presence of spin--orbit
interaction, the electronic motion must necessarily also involve
spatio-temporal dependence of the spin density; correspondingly, we
term this phenomenon \emph{spin migration}.

The study of spin migration may lead to improved control of chemical
reactivity. Of vital importance to the electronic coherence, is the
correlated motion of the nuclei. Unless there is spatial overlap of
the nuclear wavefunctions correlated with different electronic states,
coherence between the electronic states does not lead to observable
interference patterns \cite{Arnold2017,Ruberti2022}. This is
problematic for ordinary charge migration, since any attempt at
controlling chemical reactivity is significantly hampered by the
short-lived electronic coherences. The long-range Coulomb interaction
(\(\sim r^{-1}\)) will electronically couple charge fragments that are
relatively far away from one another, leading to uncontrolled
dephasing of the wavepackets. Spin migration, in contrast, fares much
better, since the spin–orbit interaction is relatively short-range
(\(\sim r^{-3}\)), which implies that once a choice has been made on
which spin-manifold the wavepacket should evolve, it will stay
there. This is a manifestation of the symmetry-protection afforded by
the spin state.

For much the same reasons, spin migration is the optimal testing
ground the most fundamental aspects of molecular spintronics
\cite{Remy2026}. Once we properly understand how the spin density
evolves in a single molecule, we can begin controlling it, for
instance using external fields such as ultrafast light pulses. This
will enable future evolutions of spintronics components such as
molecular spin valves with ultrafast controllability. We note in
passing that this field shares many aspects with the theory of
electron paramagnetic resonance (EPR) for transition metals
\cite{Autschbach2014}.


As model system, we choose to study the argon dimer cation \ch{Ar2+},
where the separation of vibronic states are comparable to the
spin--orbit splitting in the argon atoms at infinite separation. Hence,
even a simple linear combination of a few vibronic eigenstates can be
expected to induce non-trivial spatio--temporal dependence of the spin
density, on multiple time scales simultaneously.

Unless explicitly stated otherwise, we employ Hartree atomic units
(\(\hbar=e=a_0=\electronmass=1\)) throughout, as well as an extended
Einstein summation convention, where all indices appearing on one side
of an equality sign only are implicitly summed/integrated over.

\section{Theory}
We describe the system using the Born--Huang diabatized
equations. Neglecting molecular rotations, the time-dependent
Schrödinger equation (TDSE) for the nuclear wavepacket, in the absence
of external electric fields and in the centre-of-mass reference frame,
is given by
\begin{equation}
  \label{eqn:nuclear-tdse}
  [\imdt - \epsilon_a(R_{12})
  -\frac{\partial_{R_{12}}^2}{2\mu}]\ket*{\chi_a} = 0,
\end{equation}
where \(R_{12}\) the internuclear distance,
\(\mu = M_{\ch{Ar}}/2 \approx \SI{19.98}{u}\) is the reduced mass of the
nuclei, \(\epsilon_a(R_{12})\) is the potential energy curve (PEC)
corresponding to the electronic state \(i\), and \(\ket*{\chi_a}\) is the
nuclear wavefunction correlated to this electronic state. Since there
is no external electric field present, each nuclear wavefunction
\(\ket{\chi_a}\) evolves independently from the other
wavefunctions. 

\subsubsection{Spin-resolved electronic reduced density matrices}
The electronic density matrix is given by
\begin{equation}
  \label{eqn:density-matrix}
  \densitymatrix(x,x';X) =
  \ketbra{\psi_a(x;X)}{\psi_b(x';X)}.
\end{equation}
The spin-resolved transition reduced density matrix (TRDM) between the
electronic states \(a\) and \(b\) is given by
\begin{equation}
  \gamma_{ab}^{\sigma_i\sigma_j}(\vec{r},\vec{r}';X) =
  N
  \int\diffexc{1}
  \matrixel*{a\sigma_i}{\densitymatrix(x,x';X)}{b\sigma_j},
\end{equation}
where \(N\) is the number of electrons, and \(\diffexc{1}\) denotes
integration over all electronic coordinates except the first one
\cite[cf. Eq.~(17) of]{Loewdin1955a}. This remaining coordinate we
label \(\vec{r}\). From the TRDM, we can compute the spin-resolved
electronic coherence by additionally tracing out the nuclear
wavefunctions:
\begin{equation}
  \label{eqn:electronic-coherence-spin-density}
  \gamma_{ab}^{\sigma_i\sigma_j}(\vec{r},\vec{r}') =
  \int\diff{X}
  \chi_a(X)
  \gamma_{ab}^{\sigma_i\sigma_j}(\vec{r},\vec{r}';X)
  \conj{\chi_b}(X).
\end{equation}
The electronic coherence spin density is then given by
\(\gamma_{ab}^{\sigma_i\sigma_j}(\vec{r},\vec{r})\), i.e.\ along the diagonal
\(\vec{r}'=\vec{r}\). The transition amplitude associated with any
one-electron operator \(\operator{d}\) can similarly be computed as
\begin{equation}
  \label{eqn:transition-amplitude}
  \matrixel*{a\sigma_i}{\operator{d}}{b\sigma_j} =
  \int\diff{\vec{r}}
  \operator{d}
  \gamma_{ab}^{\sigma_i\sigma_j}(\vec{r},\vec{r}'),
\end{equation}
where the operator acts on the unprimed coordinate \(\vec{r}\), and
subsequently \(\vec{r}'\) is set equal to \(\vec{r}\) before carrying
out the integration \cite{Loewdin1955a}.

For the case when there is no external field, i.e.\ \(I_{ij}(t,X)=0\),
it is useful to expand \(\chi_a(X)\) in terms of the vibrational
eigenmodes of the electronic state \(a\),
\begin{equation}
  \label{eqn:vibrational-expansion}
  \chi_a(X,t) =
  c_{a_p} v_{a_p}(X,t) \equiv
  c_{a_p} \ce^{-\im E_{a_p}t}
  v_{a_p}(X,0),
\end{equation}
and tabulate \(\gamma_{a_pb_q}^{\sigma_i\sigma_j}(\vec{r},\vec{r}')\) for all pairs
of vibrational state of interest. Then the time-dependent spin density
can simply reconstructed by a contraction with the expansion
coefficients whose time dependence is given by the trivial phase
evolution:
\begin{equation}
  \label{eqn:electronic-coherence-spin-density-eigenstate-resolved}
  \begin{aligned}
    \gamma_{ab}^{\sigma_i\sigma_j}(\vec{r},\vec{r}')
    &=
      c_{a_p} \ce^{-\im E_{a_p}t}
      \gamma_{a_pb_q}^{\sigma_i\sigma_j}(\vec{r},\vec{r}')
      \conj{c_{b_q}} \ce^{\im E_{b_q}t} \\
    &=
      c_{a_p}\conj{c_{b_q}}
      \ce^{-\im (E_{a_p}-E_{b_q})t}
      \gamma_{a_pb_q}^{\sigma_i\sigma_j}(\vec{r},\vec{r}') \\
    &\defd
      \rho_{a_pb_q}
      \ce^{-\im E_{a_pb_q}t}
      \gamma_{a_pb_q}^{\sigma_i\sigma_j}(\vec{r},\vec{r}').
  \end{aligned}
\end{equation}
This approach is of course only viable when relatively few vibrational
states are required, and disassociation can be neglected.

We may now compute the differential spin density as
\begin{equation}
  \tilde{S}(\vec{r}) =
  \gamma_{ab}^{\alpha\alpha}(\vec{r},\vec{r}) -
  \gamma_{ab}^{\beta\beta}(\vec{r},\vec{r}),
\end{equation}
where \(\gamma_{ab}^{\sigma_i\sigma_j}(\vec{r},\vec{r}')\) is the spin-resolved
transition reduced density matrix, obtained by tracing out the nuclear
wavepacket and all electrons but one (see
appendix~\ref{app:interpolative-integration}).

\subsection{Initial Electron Localization}

A common model for the initial state in charge migration is the sudden
approximation, where one electron is removed from an atomic site, and
the remaining hole is initially localized to that site. As time
evolves, the charge density will redistribute over the
molecule. However, since the molecule is typically smaller than the
wavelength of the ionizing radiation, such an initial state is
difficult to prepare using weak-field ionization in a homonuclear
diatomic. In heteronuclear diatomics, with naturally asymmetry
molecular orbital, this becomes simpler.

\subsubsection{Optimized Wavepacket}
Here we describe how to construct an initial state which is a
variation of the sudden approximation idea with localization; namely
how to tailor the initial nuclear wavepacket such that the
corresponding electronic density obeys a prescribed function
\(f(\vec{r})\). This can be achieved in an optimal control fashion,
where a pulse sequence generates the desired outcome. We can formulate
this as a minimization problem:
\begin{equation}
  \begin{aligned}
    \epsilon>
    &\min_\chi
      \norm{f(\vec{r}) - \gamma_{ab}^{\sigma_i\sigma_j}(\vec{r},\vec{r})}{2} =\\
    &\min_\chi
      \norm{f(\vec{r}) -
      \int\diff{X}
      \chi_a(X)
      \gamma_{ab}^{\sigma_i\sigma_j}(\vec{r},\vec{r};X)
      \conj{\chi_b}(X)
      }{2}.
  \end{aligned}
\end{equation}
If we require \(\Real\ni f(\vec{r}) \ge 0\), then we must have \(a=b\) and
\(\sigma_i=\sigma_j\), but in the general case \(f(\vec{r})\in\Complex\), we can
prescribe an initial \emph{coherence} instead of an initial density.

We may additionally require an overall shape of the initial nuclear
wavepacket:
\begin{equation}
  \label{eqn:wavepacket-shape}
  \chi_a(X) = g_ag(X),
\end{equation}
e.g.\ in the Frank–Condon approximation we would choose \(g(X)\) to be
the vibrational ground state of the neutral. The minimization problem
is then reduced to finding a matrix \(\mat{G} = \{g_a\conj{g_b}\}\)
such that
\begin{equation}
  \epsilon>
  \min_{\mat{G}}
  \norm{f(\vec{r}) -
    g_a\conj{g_b}
    \int\diff{X}
    g^2(X)
    \gamma_{ab}^{\sigma_i\sigma_j}(\vec{r},\vec{r};X)
  }{2}.
\end{equation}
Again, if we require \(f(\vec{r})\in\Real\), \(\mat{G}\) has to be
real-symmetric, and should be chosen such that \(\det(\mat{G})=1\) if
a normalized initial state is desired.

Finally, if the nuclear wavepacket is expanded over a limited set of
vibrational eigenmodes as in Eq.~\eqref{eqn:vibrational-expansion}, we
will not be able to achieve the desired wavepacket shape
Eq.~\eqref{eqn:wavepacket-shape} identically. Instead, we will have
the best approximation in the least-square sense
\begin{equation}
  \chi_a(X) = g_a c_{a_p} \nu_{a_p},
\end{equation}
where the vibrational expansion coefficients are computed as
\begin{equation}
  c_{a_p} =
  \int\diff{X}
  \conj{\nu_{a_p}}(X)g(X).
\end{equation}
The sum of all expansion coefficients should be unity:
\begin{equation}
  \sum_p \abs*{c_{a_p}}^2 = 1;
\end{equation}
if not enough vibrational states are used, the coefficients should be
renormalized.

\subsubsection{Sudden Removal of a Molecular Orbital}
An alternative approach, that would more closely correspond to a
weak-field ionization event realizable using e.g.\ a synchrotron light
source, is the sudden removal of a molecular orbital, which is
delocalized over the whole molecule, in contrast to the atomic
orbitals. To achieve this, we first compute the electronic Dyson
orbital
\begin{equation}
  \label{eqn:dyson-orbital}
  \ket*{\psi_a^D(x;X)} \defd
  \sqrt{N}
  \braket{\psi_a(x;X)}{\psi_0(x;X)},
\end{equation}
where \(\ket{\psi_0(x;X)}\) is the electronic ground state of the neutral
with \(N\) electrons. Since the Dyson orbital is expanded in terms of
the MOs of the neutral,
\begin{equation}
  \ket*{\psi_a^D(x;X)} \equiv
  d_{ak}(X)
  \ket{\phi_k(x;X)},
\end{equation}
the (complex) amplitude of initial nuclear wavepacket associated with
the electronic state \(\ket{\psi_a}\) upon removal of the MO
\(\ket{\phi_k}\) is simply
\begin{equation}
  \label{eqn:dyson-orbital-MO-removal}
  \chi_a(X) =
  d_{ak}(X)
  \chi_0(X),
\end{equation}
where \(\chi_0(X)\) is the vibrational ground state of the neutral.

\subsection{Computational Details Specific to the Argon Dimer}

\begin{table}[htb]
  \centering
  \begin{tabular}{lllll}
    State \#
    &
      1 & 2 & 3 & 4 \\
    Label
    &
      \(^{2}\Sigma_{1/2}^u\)
        &
          \(^{2}\Sigma_{-1/2}^u\)
            &
              \(^{2}\Sigma_{-1/2}^g\)
                &
                  \(^{2}\Sigma_{1/2}^g\)
                  \Bstrut
    \\\hline
    \Tstrut
    &
      5 & 6 & 7 & 8 \\
    &
      \(^{2}\Pi_{3/2}^g\)
        &
          \(^{2}\Pi_{-3/2}^g\)
            &
              \(^{2}\Pi_{-1/2}^u\)
                &
                  \(^{2}\Pi_{1/2}^u\)
                  \Bstrut
    \\\hline
    \Tstrut
    &
      9 & 10 & 11 & 12 \\
    &
      \(^{2}\Pi_{3/2}^u\)
        &
          \(^{2}\Pi_{-3/2}^u\)
            &
              \(^{2}\Pi_{1/2}^g\)
                &
                  \(^{2}\Pi_{-1/2}^g\)
  \end{tabular}
  \caption{Labelling of electronic states of the argon dimer cation
    according to their dominant configurations}
  \label{tab:electronic-states}
\end{table}

\begin{table*}[htb]
  \caption{\label{tab:partial-populations} Contribution (populations) to different
    excited states of the cation, upon sudden removal of a molecular
    orbital (MO) from the neutral. The orbital energies \(\epsilon\) and
    vertical ionization potentials \(\ionpotential\) are given at the
    equilibrium separation \(R_{12}=\SI{3.8}{\angstrom}\). The reason
    that the cationic populations do not add up to unity, is that a
    non-negligible fraction of disassociative continuum states are
    populated upon sudden removal of a neutral MO.}
  \myruledtabular{0.86\linewidth}{rS[table-format=3.4]|S[table-format=1.4]S[table-format=1.4]S[table-format=1.4]S[table-format=1.4]S[table-format=1.4]S[table-format=1.4]S[table-format=1.4]S[table-format=1.4]S[table-format=1.4]S[table-format=1.4]S[table-format=1.4]S[table-format=1.4]|S[table-format=1.4]}{
    &  & \textbf{$^{2}\Sigma_{1/2}^u$} & \textbf{$^{2}\Sigma_{-1/2}^u$} & \textbf{$^{2}\Sigma_{-1/2}^g$} & \textbf{$^{2}\Sigma_{1/2}^g$} & \textbf{$^{2}\Pi_{3/2}^g$} & \textbf{$^{2}\Pi_{-3/2}^g$} & \textbf{$^{2}\Pi_{-1/2}^u$} & \textbf{$^{2}\Pi_{1/2}^u$} & \textbf{$^{2}\Pi_{3/2}^u$} & \textbf{$^{2}\Pi_{-3/2}^u$} & \textbf{$^{2}\Pi_{1/2}^g$} & \textbf{$^{2}\Pi_{-1/2}^g$} & \textbf{$\sum$} \\
    & $\epsilon~\backslash~\ionpotential$ [Ha] & 0.5534 & 0.5534 & 0.5697 & 0.5697 &
    0.5586 & 0.5586 & 0.5655 & 0.5655 & 0.5606 & 0.5606 & 0.5615 &
    0.5615 & \\\hline
    \(a_{g}\alpha\) & -0.5977 &  &  & 0.8100 &  &  &  &  &  &  &  &  & 0.1721 & 0.9821 \\
    \(a_{g}\beta\) & -0.5977 &  &  &  & 0.8100 &  &  &  &  &  &  & 0.1721 &  & 0.9821 \\
    \(b_{2u}\alpha\) & -0.5915 & 0.0360 &  &  &  &  &  &  & 0.4551 &  & 0.4911 &  &  & 0.9822 \\
    \(b_{2u}\beta\) & -0.5915 &  & 0.0360 &  &  &  &  & 0.4551 &  & 0.4911 &  &  &  & 0.9822 \\
    \(b_{3u}\alpha\) & -0.5915 & 0.0360 &  &  &  &  &  &  & 0.4551 &  & 0.4911 &  &  & 0.9822 \\
    \(b_{3u}\beta\) & -0.5915 &  & 0.0360 &  &  &  &  & 0.4551 &  & 0.4911 &  &  &  & 0.9822 \\
    \(b_{2g}\alpha\) & -0.5895 &  &  &  & 0.0860 &  & 0.4911 &  &  &  &  & 0.4050 &  & 0.9821 \\
    \(b_{2g}\beta\) & -0.5895 &  &  & 0.0860 &  & 0.4911 &  &  &  &  &  &  & 0.4050 & 0.9821 \\
    \(b_{3g}\alpha\) & -0.5895 &  &  &  & 0.0860 &  & 0.4911 &  &  &  &  & 0.4050 &  & 0.9821 \\
    \(b_{3g}\beta\) & -0.5895 &  &  & 0.0860 &  & 0.4911 &  &  &  &  &  &  & 0.4050 & 0.9821 \\
    \(b_{1u}\alpha\) & -0.5836 &  & 0.9099 &  &  &  &  & 0.0720 &  &  &  &  &  & 0.9819 \\
    \(b_{1u}\beta\) & -0.5836 & 0.9099 & & & & & & & 0.0720 & & & & &
    0.9819 }
\end{table*}

For a diatomic molecule, the nuclear hypercoordinate \(X\) is replaced
by the internuclear separation \(R_{12}\), and the nuclear masses by
the reduced mass
\(\mu = M_{\ch{Ar}}M_{\ch{Ar}}/(M_{\ch{Ar}}+M_{\ch{Ar}}) = M_{\ch{Ar}}/2
\approx \SI{19.98}{u}\). Since we neglect molecular rotations, we are in
either Hund's case (a) or (c), or in an intermediate regime between
the two, depending on the nuclear separation.

\begin{figure}[htb]
  \centering
  \includegraphics[width=0.99\linewidth]{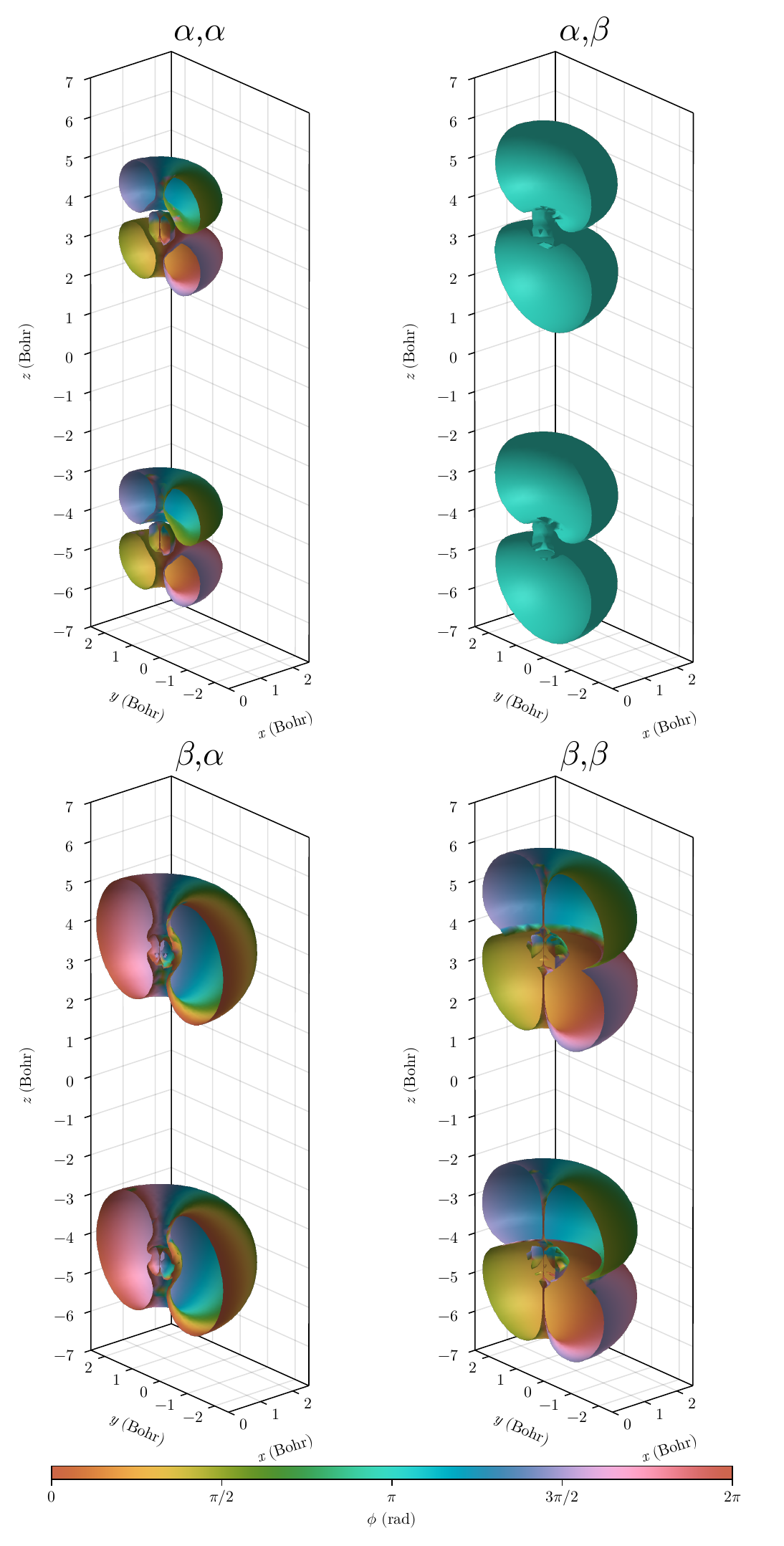}
  \caption{Example of the spin-resolved TRDMs between the states
    \(^2\Sigma^u_{1/2}\) and \(^2\Pi^u_{-1/2}\) for the internuclear distance
    \(R_{12}=\SI{7.94}{Bohr}\). The isosurfaces were computed for the
    magnitude
    \(\abs*{\gamma_{ab}^{\sigma_i\sigma_j}(\vec{r},\vec{r};R)}=\num{e-3}\), and the
    phase is indicated by the colour.}
  \label{fig:spin-resolved-trdm}
\end{figure}

The electronic states are labelled by their dominant configurations,
according to Table~\ref{tab:electronic-states}. These states and their
energies leading to the potential energy curves are computed using
GAMESS-US \cite{Schmidt1993} electronic structure program, employing
an aug-cc-pCVTZ-DK basis. The first 12 electronic states were found
using a full-valence CAS (11e,6o) for the scalar-relativistic
DK-CASSCF calculation. Then spin--orbit coupling was introduced using a
SO-MCQDPT2 calculation, were the orbitals were kept fixed, affecting
only the CI coefficients. From these wavefunctions, the spin-resolved
TRDMs were computed using superdyson \cite{superdyson}:
\begin{equation}
  \gamma_{ab}^{\sigma_i\sigma_j}(\vec{r},\vec{r}';R) =
  C_{\mu a}(R)
  \gamma_{\mu\nu}^{\sigma_i\sigma_j}(\vec{r},\vec{r}';R)
  \conj{C_{\nu b}}(R),
\end{equation}
where \(C_{\mu a}(R)\) are the expansion coefficients of the electronic
states in terms of bosonic irreps multiplied by spin functions. For
each pair of these basis functions, and each combination of spins
\(\sigma_i,\sigma_j\in\{\spinup,\spindown\}\), the density matrix
\(\gamma_{\mu\nu}^{\sigma_i\sigma_j}(\vec{r},\vec{r}';R)\) is encoded as a singular value
decomposition where each pair left- and right-singular vectors contain
expansion coefficients in the underlying MO basis
\(\{\phi(\vec{r};R)\}\), which depends on the electronic coordinate
\(\vec{r}\) and \emph{parametrically} on the distance \(R\) from the
centre of mass. An example of a spin-resolved TRDM is shown in
Fig.~\ref{fig:spin-resolved-trdm}.

\subsubsection{Numerical Solution of the Vibrational Problem}

\begin{figure}[htb]
  \centering
  \includegraphics[trim=30cm 30cm 30cm 30cm]{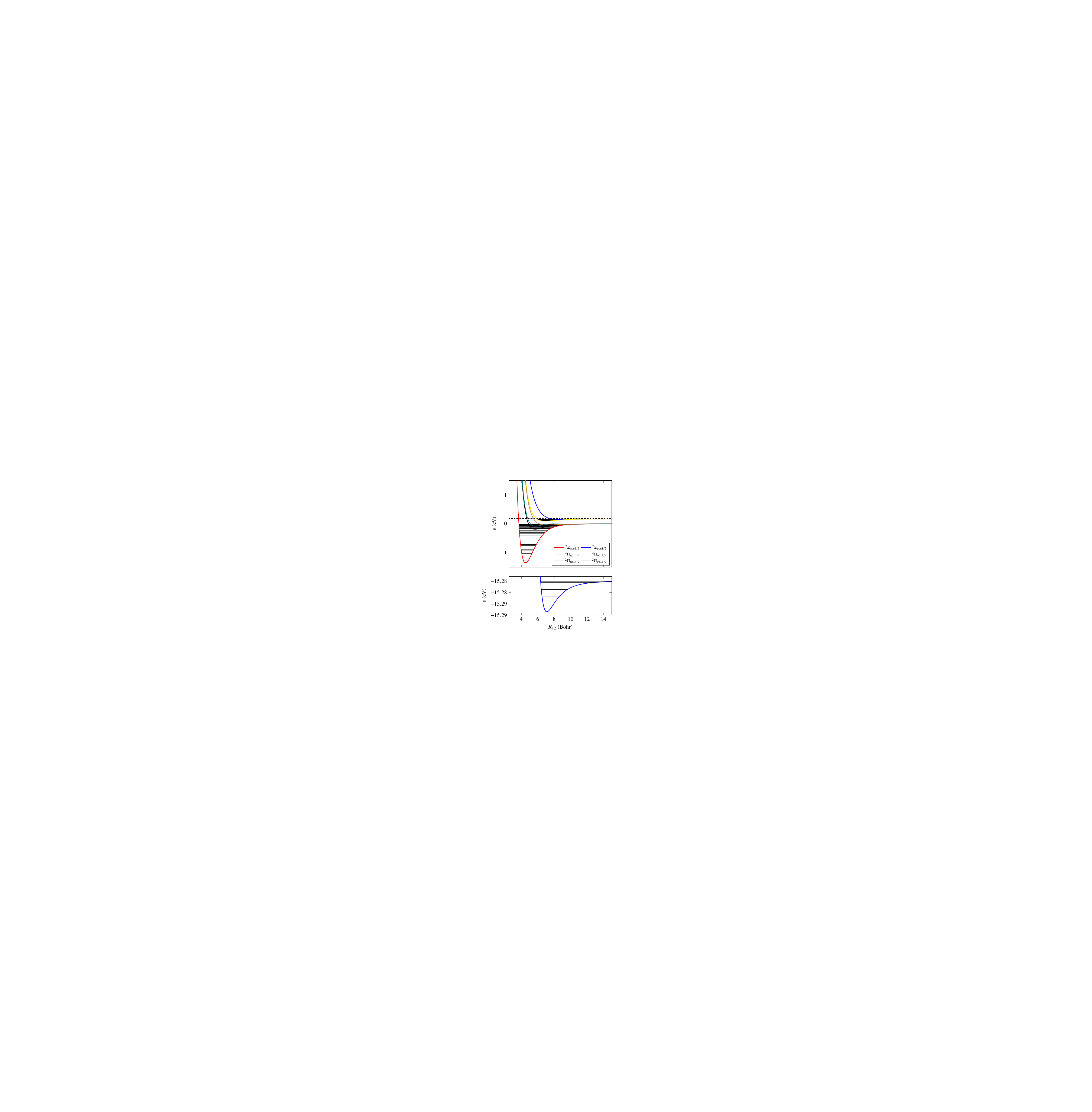}
  \caption{Potential energy curves (PECs) for the argon dimer cation
    (top panel) and neutral (bottom panel) as a function of the
    internuclear separation \(R_{12}\). Because of the time reversal
    symmetry, \(^2\Lambda_{\pi;M_\Omega}\) and
    \(^2\Lambda_{\pi;-M_\Omega}\) states (where \(\pi\) denotes parity of the state
    under inversion) are energetically degenerate, and thus share
    PEC. The dashed horizontal line indicates the position of the
    \(^2P_{1/2}\) state of the atomic cation at
    \(\sim\SI{0.178}{\electronvolt}\).}
  \label{fig:cation-pecs}
\end{figure}

Having obtained the potential energy curves \(\epsilon_i(R_{12})\) (see
Fig.~\ref{fig:cation-pecs}), we solve the vibrational problem
\eqref{eqn:nuclear-tdse}, which consists of a set of coupled
one-dimensional ones, in a finite-element discrete-variable
representation (FEDVR) \cite{Rescigno2000} basis. The grid points were
distributed according to
\begin{equation}
  R_j = R_{j-1} +
  \rho_{\textrm{min}} +
  (1-\mathrm{e}^{-\alpha R_{j-1}})
  (\rho_{\textrm{max}} - \rho_{\textrm{min}}),
\end{equation}
where \(\rho_{\textrm{max}}\approx\SI{0.18}{Bohr}\),
\(\rho_{\textrm{min}} = 10^{-2} \rho_{\textrm{max}}\), and
\(\alpha=0.1\), which leads to an approximately log--linear behaviour. The
first point of the grid is chosen as
\(R_{12,\textrm{min}}=\SI{0}{Bohr}\), and the tabulated potential
energy curves are extrapolated quadratically to avoid reflections off
of a hard boundary at \(R_{12}=0\).
Using 385 points, the grid extends to approximately
\SI{22.5}{Bohr}. The polynomial order of the DVR functions was chosen
as \(7\).

To avoid unphysical reflections at the largest \(R_{12}\), we employ
the complex-absorbing potential (CAP) by \citet{Manolopoulos2002},
with a design parameter of \(\delta=0.20866\) and the requirement that all
kinetic energies larger than \SI{e-3}{\electronvolt} are reflected
less than \SI{1}{\percent}. For particle of mass
\(\mu = M_{\ch{Ar}}/2\), the corresponding de~Broglie wavelength is
\SI{3.84}{Bohr}, which is the extent of the CAP at the end of the
computational box.

\subsubsection{Sudden Approximation}
In table~\ref{tab:partial-populations}, the partial contributions to
the various excited electronic states of the cation are given, for
sudden removal of the different molecular orbitals of the neutral.

\begin{figure}[htb]
  \centering
  \includegraphics[trim=30cm 30cm 30cm 30cm]{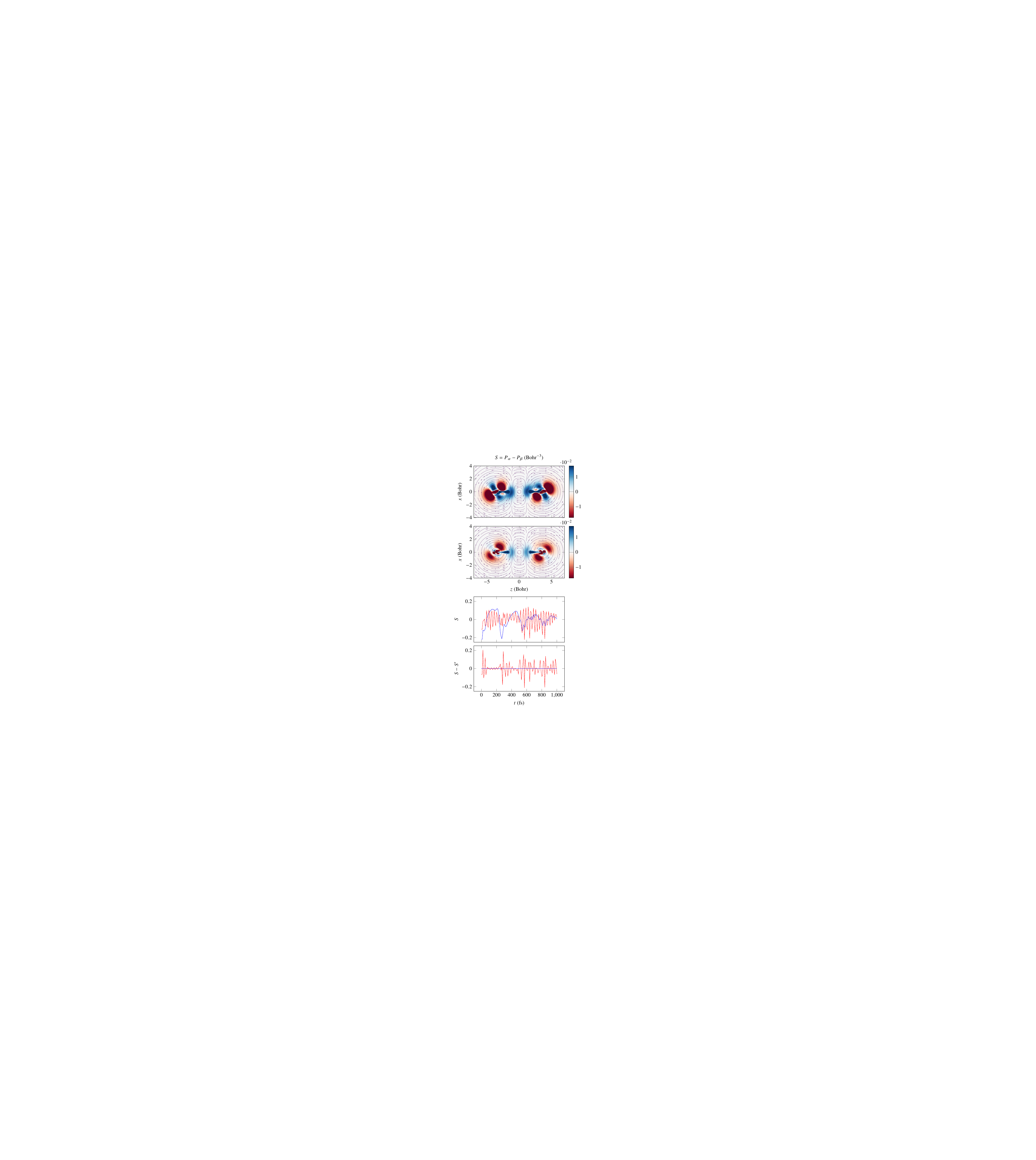}
  \caption{Sudden ionization. \emph{First and second panels}:
    Elecronic spin density in the plane \(y=\SI{0}{Bohr}\), at times
    \(t=\SI{0}{\femto\second}\), and \(t=\SI{330}{\femto\second}\),
    respectively. The spin density average over the whole time
    interval up to \SI{1000}{\femto\second} has been subtracted, to
    enhance the emphasize the relative rotary motion of the spin
    density, as is also indicated by the superposed streamlines of the
    current density. \emph{Third panel}: Line-outs at
    \((z,x)=(\SI{3.59}{Bohr},\SI{0}{Bohr})\) (blue line) and
    \((z,x)=(\SI{3.59}{Bohr},\SI{0.15}{Bohr})\) (red
    line). \emph{Fourth panel}: Difference between line-outs at
    \((z,x)\) and their mirror points at \((-z,x)\).}
  \label{fig:MO-sudden-ionization}
\end{figure}

\section{Results}
In figure~\ref{fig:MO-sudden-ionization}, show is the resulting spin
density upon sudden removal of an equal superposition of the molecular
orbitals \(b_{2u}\alpha\) and \(b_{1u}\alpha\). This superposition has enough
initial spin asymmetry between the left and right atomic sites, that
the spin density will oscillate back and forth.

\begin{figure}
  \centering
  \includegraphics[trim=30cm 30cm 30cm 30cm]{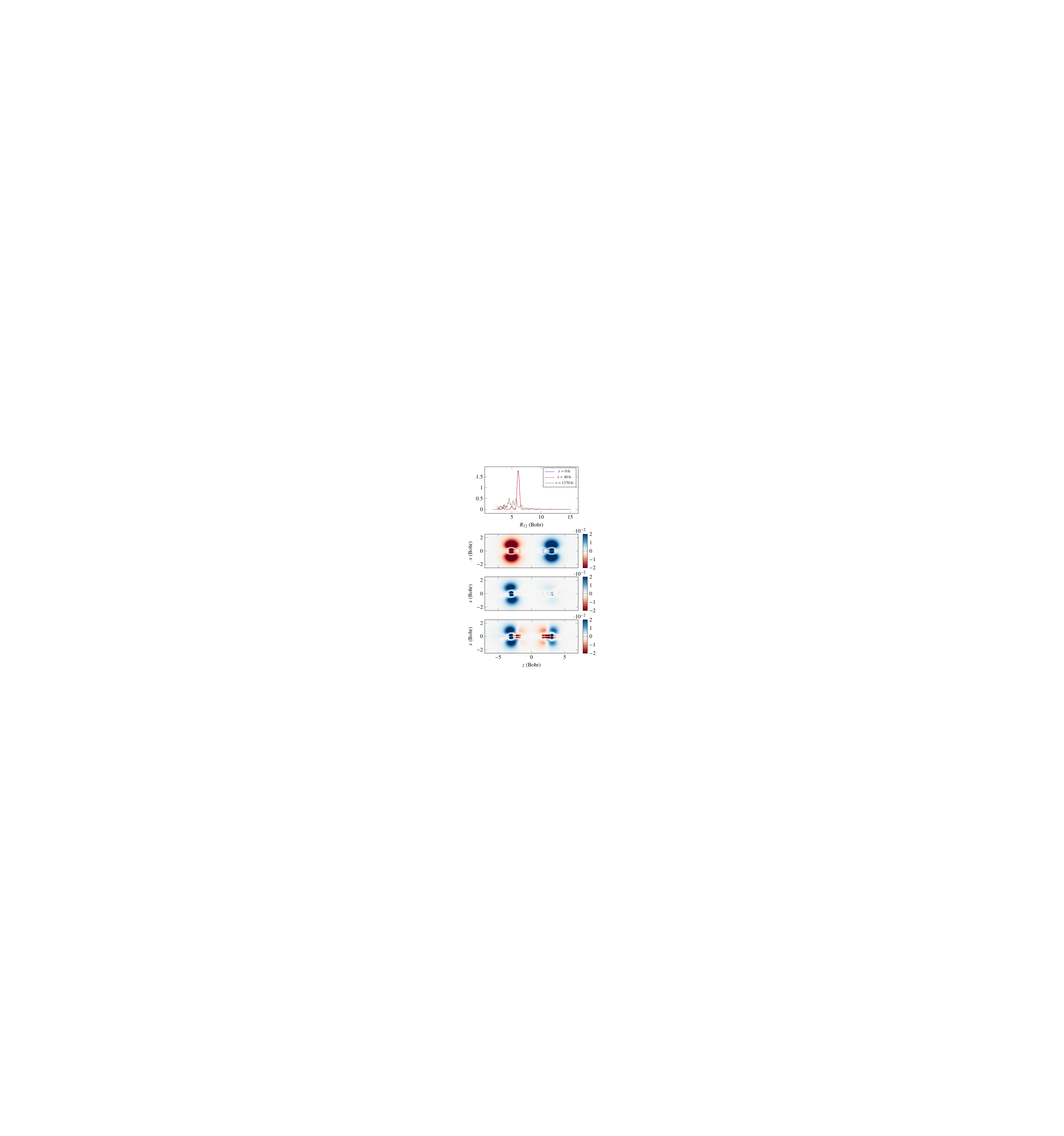}
  \caption{Evolution of the spin density, when the initial nuclear
    wavepacket has been optimized to maximize spin asymmetry at the
    starting time.}
  \label{fig:optimized-wavepacket}
\end{figure}

Perhaps more interesting is the case illustrated in
figure~\ref{fig:optimized-wavepacket}. Shown there is the wavepacket
optimized for maximal initial spin-density asymmetry between the left
and right atomic sites at the initial time. As the time progresses,
the nuclear wavepacket evolves, with a series of quasirevivals. The
resultant spin density exhibits mesoscopic motion between the two
sites, as a function of time, a behaviour that persists for very long
times (\(\ge\SI{1000}{\femto\second}\)), in spite of the nuclear
wavepacket having diffused considerably.

\begin{acknowledgments}
  MI acknowledges support from \emph{Horizon 2020 research and
    innovation} (899794).
\end{acknowledgments}

\bibliographystyle{apsrev4-2}
\bibliography{\bibliographyfile}

\begin{thebibliography}{11}%
\makeatletter
\providecommand \@ifxundefined [1]{%
 \@ifx{#1\undefined}
}%
\providecommand \@ifnum [1]{%
 \ifnum #1\expandafter \@firstoftwo
 \else \expandafter \@secondoftwo
 \fi
}%
\providecommand \@ifx [1]{%
 \ifx #1\expandafter \@firstoftwo
 \else \expandafter \@secondoftwo
 \fi
}%
\providecommand \natexlab [1]{#1}%
\providecommand \enquote  [1]{``#1''}%
\providecommand \bibnamefont  [1]{#1}%
\providecommand \bibfnamefont [1]{#1}%
\providecommand \citenamefont [1]{#1}%
\providecommand \href@noop [0]{\@secondoftwo}%
\providecommand \href [0]{\begingroup \@sanitize@url \@href}%
\providecommand \@href[1]{\@@startlink{#1}\@@href}%
\providecommand \@@href[1]{\endgroup#1\@@endlink}%
\providecommand \@sanitize@url [0]{\catcode `\\12\catcode `\$12\catcode
  `\&12\catcode `\#12\catcode `\^12\catcode `\_12\catcode `\%12\relax}%
\providecommand \@@startlink[1]{}%
\providecommand \@@endlink[0]{}%
\providecommand \url  [0]{\begingroup\@sanitize@url \@url }%
\providecommand \@url [1]{\endgroup\@href {#1}{\urlprefix }}%
\providecommand \urlprefix  [0]{URL }%
\providecommand \Eprint [0]{\href }%
\providecommand \doibase [0]{https://doi.org/}%
\providecommand \selectlanguage [0]{\@gobble}%
\providecommand \bibinfo  [0]{\@secondoftwo}%
\providecommand \bibfield  [0]{\@secondoftwo}%
\providecommand \translation [1]{[#1]}%
\providecommand \BibitemOpen [0]{}%
\providecommand \bibitemStop [0]{}%
\providecommand \bibitemNoStop [0]{.\EOS\space}%
\providecommand \EOS [0]{\spacefactor3000\relax}%
\providecommand \BibitemShut  [1]{\csname bibitem#1\endcsname}%
\let\auto@bib@innerbib\@empty
\bibitem [{\citenamefont {Breidbach}\ and\ \citenamefont
  {Cederbaum}(2003)}]{Breidbach2003}%
  \BibitemOpen
  \bibfield  {author} {\bibinfo {author} {\bibfnamefont {J.}~\bibnamefont
  {Breidbach}}\ and\ \bibinfo {author} {\bibfnamefont {L.~S.}\ \bibnamefont
  {Cederbaum}},\ }\href {https://doi.org/10.1063/1.1540618} {\bibfield
  {journal} {\bibinfo  {journal} {The Journal of Chemical Physics}\ }\textbf
  {\bibinfo {volume} {118}},\ \bibinfo {pages} {3983} (\bibinfo {year}
  {2003})}\BibitemShut {NoStop}%
\bibitem [{\citenamefont {Kuleff}\ \emph {et~al.}(2005)\citenamefont {Kuleff},
  \citenamefont {Breidbach},\ and\ \citenamefont {Cederbaum}}]{Kuleff2005}%
  \BibitemOpen
  \bibfield  {author} {\bibinfo {author} {\bibfnamefont {A.~I.}\ \bibnamefont
  {Kuleff}}, \bibinfo {author} {\bibfnamefont {J.}~\bibnamefont {Breidbach}},\
  and\ \bibinfo {author} {\bibfnamefont {L.~S.}\ \bibnamefont {Cederbaum}},\
  }\href {https://doi.org/10.1063/1.1961341} {\bibfield  {journal} {\bibinfo
  {journal} {The Journal of Chemical Physics}\ }\textbf {\bibinfo {volume}
  {123}},\ \bibinfo {pages} {044111} (\bibinfo {year} {2005})}\BibitemShut
  {NoStop}%
\bibitem [{\citenamefont {Arnold}\ \emph {et~al.}(2017)\citenamefont {Arnold},
  \citenamefont {Vendrell},\ and\ \citenamefont {Santra}}]{Arnold2017}%
  \BibitemOpen
  \bibfield  {author} {\bibinfo {author} {\bibfnamefont {C.}~\bibnamefont
  {Arnold}}, \bibinfo {author} {\bibfnamefont {O.}~\bibnamefont {Vendrell}},\
  and\ \bibinfo {author} {\bibfnamefont {R.}~\bibnamefont {Santra}},\ }\href
  {https://doi.org/10.1103/physreva.95.033425} {\bibfield  {journal} {\bibinfo
  {journal} {Physical Review A}\ }\textbf {\bibinfo {volume} {95}},\ \bibinfo
  {pages} {033425} (\bibinfo {year} {2017})}\BibitemShut {NoStop}%
\bibitem [{\citenamefont {Ruberti}\ \emph {et~al.}(2022)\citenamefont
  {Ruberti}, \citenamefont {Patchkovskii},\ and\ \citenamefont
  {Averbukh}}]{Ruberti2022}%
  \BibitemOpen
  \bibfield  {author} {\bibinfo {author} {\bibfnamefont {M.}~\bibnamefont
  {Ruberti}}, \bibinfo {author} {\bibfnamefont {S.}~\bibnamefont
  {Patchkovskii}},\ and\ \bibinfo {author} {\bibfnamefont {V.}~\bibnamefont
  {Averbukh}},\ }\href {https://doi.org/10.1039/d2cp01562e} {\bibfield
  {journal} {\bibinfo  {journal} {Physical Chemistry Chemical Physics}\
  }\textbf {\bibinfo {volume} {24}},\ \bibinfo {pages} {19673} (\bibinfo {year}
  {2022})}\BibitemShut {NoStop}%
\bibitem [{\citenamefont {Remy}\ and\ \citenamefont {Mangin}(2026)}]{Remy2026}%
  \BibitemOpen
  \bibfield  {author} {\bibinfo {author} {\bibfnamefont {Q.}~\bibnamefont
  {Remy}}\ and\ \bibinfo {author} {\bibfnamefont {S.}~\bibnamefont {Mangin}},\
  }\href {https://doi.org/10.1016/j.jmmm.2026.174344} {\bibfield  {journal}
  {\bibinfo  {journal} {Journal of Magnetism and Magnetic Materials}\ }\textbf
  {\bibinfo {volume} {655}},\ \bibinfo {pages} {174344} (\bibinfo {year}
  {2026})}\BibitemShut {NoStop}%
\bibitem [{\citenamefont {Autschbach}(2014)}]{Autschbach2014}%
  \BibitemOpen
  \bibfield  {author} {\bibinfo {author} {\bibfnamefont {J.}~\bibnamefont
  {Autschbach}},\ }\href {https://doi.org/10.1098/rsta.2012.0489} {\bibfield
  {journal} {\bibinfo  {journal} {Philosophical Transactions of the Royal
  Society A: Mathematical, Physical and Engineering Sciences}\ }\textbf
  {\bibinfo {volume} {372}},\ \bibinfo {pages} {20120489} (\bibinfo {year}
  {2014})}\BibitemShut {NoStop}%
\bibitem [{\citenamefont {L{\"o}wdin}(1955)}]{Loewdin1955a}%
  \BibitemOpen
  \bibfield  {author} {\bibinfo {author} {\bibfnamefont {P.-O.}\ \bibnamefont
  {L{\"o}wdin}},\ }\href {https://doi.org/10.1103/physrev.97.1474} {\bibfield
  {journal} {\bibinfo  {journal} {Physical Review}\ }\textbf {\bibinfo {volume}
  {97}},\ \bibinfo {pages} {1474} (\bibinfo {year} {1955})}\BibitemShut
  {NoStop}%
\bibitem [{\citenamefont {Schmidt}\ \emph {et~al.}(1993)\citenamefont
  {Schmidt}, \citenamefont {Baldridge}, \citenamefont {Boatz}, \citenamefont
  {Elbert}, \citenamefont {Gordon}, \citenamefont {Jensen}, \citenamefont
  {Koseki}, \citenamefont {Matsunaga}, \citenamefont {Nguyen}, \citenamefont
  {Su}, \citenamefont {Windus}, \citenamefont {Dupuis},\ and\ \citenamefont
  {Montgomery}}]{Schmidt1993}%
  \BibitemOpen
  \bibfield  {author} {\bibinfo {author} {\bibfnamefont {M.~W.}\ \bibnamefont
  {Schmidt}}, \bibinfo {author} {\bibfnamefont {K.~K.}\ \bibnamefont
  {Baldridge}}, \bibinfo {author} {\bibfnamefont {J.~A.}\ \bibnamefont
  {Boatz}}, \bibinfo {author} {\bibfnamefont {S.~T.}\ \bibnamefont {Elbert}},
  \bibinfo {author} {\bibfnamefont {M.~S.}\ \bibnamefont {Gordon}}, \bibinfo
  {author} {\bibfnamefont {J.~H.}\ \bibnamefont {Jensen}}, \bibinfo {author}
  {\bibfnamefont {S.}~\bibnamefont {Koseki}}, \bibinfo {author} {\bibfnamefont
  {N.}~\bibnamefont {Matsunaga}}, \bibinfo {author} {\bibfnamefont {K.~A.}\
  \bibnamefont {Nguyen}}, \bibinfo {author} {\bibfnamefont {S.}~\bibnamefont
  {Su}}, \bibinfo {author} {\bibfnamefont {T.~L.}\ \bibnamefont {Windus}},
  \bibinfo {author} {\bibfnamefont {M.}~\bibnamefont {Dupuis}},\ and\ \bibinfo
  {author} {\bibfnamefont {J.~A.}\ \bibnamefont {Montgomery}},\ }\href
  {https://doi.org/10.1002/jcc.540141112} {\bibfield  {journal} {\bibinfo
  {journal} {Journal of Computational Chemistry}\ }\textbf {\bibinfo {volume}
  {14}},\ \bibinfo {pages} {1347} (\bibinfo {year} {1993})}\BibitemShut
  {NoStop}%
\bibitem [{\citenamefont {Patchkovskii}(2025)}]{superdyson}%
  \BibitemOpen
  \bibfield  {author} {\bibinfo {author} {\bibfnamefont {S.}~\bibnamefont
  {Patchkovskii}},\ }\href {https://github.com/MBI-Theory/superdyson} {\bibinfo
  {title} {Superdyson}} (\bibinfo {year} {2025})\BibitemShut {NoStop}%
\bibitem [{\citenamefont {Rescigno}\ and\ \citenamefont
  {McCurdy}(2000)}]{Rescigno2000}%
  \BibitemOpen
  \bibfield  {author} {\bibinfo {author} {\bibfnamefont {T.~N.}\ \bibnamefont
  {Rescigno}}\ and\ \bibinfo {author} {\bibfnamefont {C.~W.}\ \bibnamefont
  {McCurdy}},\ }\href {https://doi.org/10.1103/physreva.62.032706} {\bibfield
  {journal} {\bibinfo  {journal} {Physical Review A}\ }\textbf {\bibinfo
  {volume} {62}},\ \bibinfo {pages} {032706} (\bibinfo {year}
  {2000})}\BibitemShut {NoStop}%
\bibitem [{\citenamefont {Manolopoulos}(2002)}]{Manolopoulos2002}%
  \BibitemOpen
  \bibfield  {author} {\bibinfo {author} {\bibfnamefont {D.~E.}\ \bibnamefont
  {Manolopoulos}},\ }\href {https://doi.org/10.1063/1.1517042} {\bibfield
  {journal} {\bibinfo  {journal} {J. Chem. Phys.}\ }\textbf {\bibinfo {volume}
  {117}},\ \bibinfo {pages} {9552} (\bibinfo {year} {2002})}\BibitemShut
  {NoStop}%
\end{thebibliography}%

\appendix

\section{Interpolative Integration}
\label{app:interpolative-integration}
In equation~\eqref{eqn:electronic-coherence-spin-density} [or
equivalently
equation~\eqref{eqn:electronic-coherence-spin-density-eigenstate-resolved}],
we need to integrate over the nuclear coordinates \(X\). Typically, we
need a denser grid in \(X\) to accurately represent the vibrational
functions \(\chi_a(X)\), than is available for the tabulation of the
various quantities; \(C_{\mu a}(X)\),
\(\gamma_{\mu\nu}^{\sigma_i\sigma_j}(\vec{r},\vec{r}';X)\), and the molecular orbitals
(MOs). We therefore need to interpolate these latter quantities onto
the quadrature grid on which the functions \(\chi_a(X)\) are resolved.

Without loss of generality, we may consider a one-dimensional
integral:
\begin{equation}
  I =
  \int\diff{x}
  f(x)g(x) \approx
  w_i f(x_i) g(x_i)
  \defd w_i f_i g_i,
\end{equation}
where in the second step, we approximated the integral using a
suitable quadrature rule, with quadrature nodes \(\{x_i\}\) and
corresponding weights \(\{w_i\}\). Now, assume we know \(g(x)\) only
on a coarser, not necessarily regular, grid \(\{\tilde{x}_j\}\):
\(\tilde{g}_j\defd g(\tilde{x}_j)\). Our task is then to find an
interpolation that yields an approximation of \(g(x)\) on our desired,
finer grid: \(g_i = u_{ij}\tilde{g}_j\). We can then write the
integral as
\begin{equation}
  I \approx (w_if_iu_{ij})\tilde{g}_j
  \defd
  \tilde{f}_j\tilde{g}_j,
\end{equation}
where we use the parentheses to indicate the preferred order of
contraction. Since additionally, the evaluation of \(\tilde{g}_j\) may
be costly, it is sensible to generate a permutation vector \(\pi\) such
that \(\abs{\tilde{f}_{\pi_{n-1}}} > \abs{\tilde{f}_{\pi_n}}\), and
compute the integral as
\begin{equation}
  I \approx I_{\pi_{n-1}} + \delta{}I_{\pi_n}, \quad
  \delta{}I_{\pi_n} \defd
  \tilde{f}_{\pi_n}\tilde{g}_{\pi_n},
\end{equation}
terminating the sequence when \(\delta{}I_{\pi_n}<\epsilon\), for some tolerance
\(\epsilon\).

\end{document}